\documentclass[letterpaper]{article} % DO NOT CHANGE THIS
\usepackage{aaai2026}  % DO NOT CHANGE THIS
\usepackage{times}  % DO NOT CHANGE THIS
\usepackage{helvet}  % DO NOT CHANGE THIS
\usepackage{courier}  % DO NOT CHANGE THIS
\usepackage[hyphens]{url}  % DO NOT CHANGE THIS
\usepackage{graphicx} % DO NOT CHANGE THIS
\usepackage{natbib}  % DO NOT CHANGE THIS AND DO NOT ADD ANY OPTIONS TO IT
\usepackage{caption} % DO NOT CHANGE THIS AND DO NOT ADD ANY OPTIONS TO IT
\usepackage{algorithm}
\usepackage{algorithmic}

\usepackage{xcolor}
\usepackage{amsmath}
\usepackage{multicol}
\usepackage{subcaption}
\usepackage{booktabs}
\usepackage{multirow}
\usepackage{amssymb}

\newtheorem{definition}{Definition}

\usepackage{newfloat}
\usepackage{listings}
\DeclareCaptionStyle{ruled}{labelfont=normalfont,labelsep=colon,strut=off} % DO NOT CHANGE THIS
\floatstyle{ruled}
\newfloat{listing}{tb}{lst}{}
\floatname{listing}{Listing}
\title{Verification-Guided Context Optimization for Tool Calling via Hierarchical LLMs-as-editors}

\author{
    Henger Li,
    Shuangjie You,
    Flavio Di Palo,
    Yiyue Qian,
    Ayush Jain
}

\affiliations{
    Amazon SPX AI Lab\\
    \{henger, sjyou, paloflav, iamyiyue, jainays\}@amazon.com
}

\usepackage{bibentry}
\begin{document}

\maketitle

\begin{abstract}
Tool calling enables Large Language Models (LLMs) to interact with external environments via tool invocation, providing a practical mechanism to overcome the inherent limitations of pretraining. However, the effectiveness of tool use depends critically on the quality of associated documentation and knowledge base context, which are typically authored for human users and often misaligned with LLMs' interpretive needs. This issue is further amplified in industrial settings, where hundreds of tools with overlapping functionalities introduce challenges of scalability, variability, and ambiguity.
We propose \textbf{Verification-Guided Context Optimization (VGCO)}, a framework that employs LLMs-as-editors to automatically refine tool-related documentation and knowledge base context. VGCO operates in two stages: (1) \textbf{Evaluation}, which collects real-world failure cases and identifies tool-context mismatches; and (2) \textbf{Optimization}, which performs hierarchical editing through offline learning with structure-aware, in-context optimization.
The novelty of our LLM editors lies in three key aspects: (1) a \textbf{hierarchical structure} that integrates naturally into the tool-calling workflow; (2) a \textbf{state-aware, action-specific, and verification-guided} design that constrains the search space for efficient, targeted improvements; and (3) the potential for \textbf{cost-efficient sub-task specialization} through either prompt engineering large editor models or post-training smaller editor models. Unlike prior work emphasizing multi-turn reasoning, VGCO targets the single-turn \textbf{large scale tool-calling problem}, achieving significant gains in accuracy, robustness, and generalization across LLMs.
\end{abstract}

\section{Introduction}
%\textcolor{blue}{HL: In both the abstract and introduction, we should emphasize our novelty and specificity in the problem setting. While prior research has primarily focused on the depth of workflows—reasoning and planning in multi-turn scenarios—our work targets the single-turn problem, which is more common and practical in real industrial applications. This focus allows us to address key challenges of scalability, variability and ambiguously when selecting from over 100 tools, many of which have overlapping functionalities.}
Large Language Models (LLMs) have demonstrated impressive capabilities across a wide range of tasks, yet their effectiveness remains constrained by the static nature of their pretraining \citep{wang2024tools, shen2024llm}. Tool calling—a mechanism that enables LLMs to invoke external tools—has emerged as a promising solution to this limitation by allowing models to access real-time information, perform complex computations, or interact with structured environments \citep{qu2025tool, qin2024tool}. However, the success of tool-augmented LLMs depends not only on tool availability but critically on the quality, interpretability, and relevance of accompanying documentation and knowledge base contexts \citep{qu2025exploration, agrawal2025gepa}. These materials are often authored for human consumption and consequently fail to align with the operational needs and input constraints of LLMs. This mismatch between tool interfaces and model expectations frequently leads to breakdowns in task execution, even when the appropriate tools are present.

%In real industrial scenarios, tool-related content introduces additional challenges due to high variety and large-scale tool inventories. 
In industrial settings, these challenges are amplified by high tool variety and large-scale inventories. Enterprises frequently manage hundreds of APIs and tools, many with overlapping functionalities or ambiguous boundaries. A practical and effective strategy to address these challenges is a multi-layered approach, adaptable to the complexity of the use case and the scale or diversity of the available toolset. At higher levels, selection leverages domain knowledge, as in this paper via knowledge base retrieval, to identify a manageable subset of relevant tools. Subsequent layers perform targeted tool invocation and parameter filling on this curated set. This hierarchical methodology reduces operational complexity, enhances efficiency, and ensures that models focus on the most pertinent tools, thereby addressing both scalability and interpretability challenges in large-scale enterprise environments.
Existing context optimization frameworks, such as Dspy \citep{khattab2023dspy}, SAMMO \citep{schnabel2024symbolic} and Promptim \citep{cui2025automatic}, exhibit limited effectiveness when applied to industrial-level scenarios. Three key limitations are particularly notable:
1. Existing optimizers operate in a stateless manner, disregarding the current status or structure of the context.
2. General-purpose optimizers lack a well-defined action space, which could otherwise constrain modifications to higher-level operations (e.g., edit, delete, add) rather than relying solely on token-level adjustments that often introduce edge cases.
3. Current optimizers do not incorporate evaluation frameworks (customized metrics, verification datasets, etc.) tailored to the specific domain.
%Addressing these gaps requires a principled approach to context engineering that develops an optimizer which is state-aware, supports a structured action space, and integrates domain-specific evaluation mechanisms, thereby achieving greater effectiveness than directly adopting existing open-source methods.
Addressing these gaps requires a principled approach to context engineering that develops an optimizer which is state-aware and supports a structured action space. Additionally, it should integrate domain-specific evaluation mechanisms, thereby achieving greater effectiveness than directly adopting existing open-source methods.

In this work, we present \textbf{Verification-Guided Context Optimization (VGCO)}, a framework designed to systematically refine tool-related documentation and knowledge base context for tool-augmented LLMs. VGCO is motivated by the observation that many failures in these systems stem not from model reasoning limitations, but from incomplete, inconsistent, or improperly structured contextual information.
To address the first challenge of \textbf{stateless context optimization}, our framework employs \textbf{LLMs-as-editors} to iteratively refine contextual inputs, maintaining awareness of the current state and hierarchical structure of the information. To handle the second challenge of \textbf{unstructured or unconstrained editing}, VGCO introduces a \textbf{structured action space}, enabling edits at a higher semantic level—such as add, modify, or delete—rather than relying solely on token-level adjustments. To tackle the third challenge of \textbf{lack of domain-specific evaluation}, VGCO integrates \textbf{verification-guided signals}, capturing mismatches, ground-truth references, and inference logic to guide effective context refinement.

The framework operates in two primary stages. In the \textbf{Evaluation} stage, structured signals are collected to identify and diagnose context failures. In the \textbf{Optimization} stage, revised tool documentation and contextual inputs are generated and integrated through hierarchical LLM editing, producing refined, actionable knowledge for subsequent tool invocation. To further enhance generalization, VGCO incorporates a \textbf{task-specific prompt engineering or post-training} with in-context learning (ICL) examples representing both positive and negative editing outcomes.
This design enables VGCO to systematically address key limitations of prior context optimization methods while providing a principled mechanism for improving reliability, interpretability, and controllability. The framework’s evaluation spans multiple tool-use scenarios, capturing the range of potential context failures and measuring the effectiveness of its iterative refinements. Due to space limit, we move our related work section to Appendix.
Our contributions include:

% Our contributions include: (1) VGCO framework (2) LLMs-as-editors to capture structure info of contexts with potential to improve by FT (and/or PO) (3) Extensive experiments: a. SOTA models (e.g., Claude 4) b. surpass the SOTA baseline (i.e., DRAFT) c. xLAM and AST align with BFLC to directly compare and add up to model enhance methods

\begin{itemize}
    \item \textbf{VGCO Framework}: We propose Verification-Guided Context Optimization, a structured, feedback-driven approach using LLMs to optimize tool documentations and knowledge base contexts, with potential to incorporate more context in refine loop.
    
    \item \textbf{Hierarchical LLM Editors}: We introduce a three-tiered optimization strategy—Retrieval, Tool, and Parameter levels—that captures the dependent structure of tool-use contexts and supports post training and/or prompt engineering.

    \item \textbf{Extensive Empirical Results}: VGCO achieves consistent improvement for tool-calling abilities for different models, outperforms the SOTA baselines like DRAFT, and integrates seamlessly with various datasets and metrics.
\end{itemize}

\section{Related Works}

\subsubsection{Tool Learning.}

Recent studies show that large language models (LLMs) can effectively employ external tools for complex tasks \citep{qu2025tool, qin2024tool}. Tool-learning approaches are typically tuning-based or tuning-free \citep{gao2024confucius}.
Tuning-based methods enhance tool use via fine-tuning on curated datasets. Systems such as xLAMs \citep{liu2024apigen, prabhakar2025apigen, zhang2024xlam}, ToolAce \citep{liu2024toolace}, and ToolBench \citep{qin2024toolllm} generate large-scale synthetic data through agentic pipelines. Building on these, Tool-MVR \citep{ma2025advancing} and DiaTool-DPO \citep{jung2025diatool} apply SFT and DPO on multi-turn trajectories, while others \citep{qian2025toolrl, wang2025otc, zhang2025nemotron} leverage RL for improved proficiency and generalization. Despite strong results, these methods depend on open-source models, large synthetic datasets, and significant compute.
Tuning-free methods, by contrast, bypass training by supplying tool-related context and few-shot demonstrations \citep{wei2022chain, hsieh2023tool, paranjape2023art, du2024anytool, shi2024learning, qu2025exploration}, relying on in-context learning to infer usage. This plug-and-play paradigm is lightweight but highly sensitive to context quality—metadata, prompts, and documentation must align with model comprehension \citep{yuan2024easytool}.
We propose a systematic context optimization framework to improve tuning-free tool use by refining metadata, prompts, and documentation. Grounding these elements in verifiable datasets enhances alignment with LLM reasoning and tool efficacy. Our tuning-free design prioritizes high-quality, verified data from users, auditors, and experts over large-scale synthetic data, yielding more reliable and adaptive tool interaction.

\subsubsection{Context Optimization.}

Research on Prompt Optimization (PO) advances through framework-based and algorithmic approaches. Frameworks such as DSPy \citep{khattab2023dspy} and SAMMO \citep{schnabel2024symbolic} modularize prompts for systematic refinement—DSPy provides compositional Python-based optimization, while SAMMO models prompts as function graphs. These shift PO from ad hoc design to reusable workflows, though industrial deployment still demands expert input, customization, and scalable orchestration.
Algorithmic PO methods automate prompt search and evolution: APE \citep{zhou2022large} employs paired LLMs, APO \citep{pryzant2023automatic} performs feedback-based edits, GRIPS \citep{prasad2022grips} applies heuristic search, and OPRO \citep{yang2023large} frames optimization as a meta-prompt problem. GEPA \citep{agrawal2025gepa} extends APO with reflective, evolutionary search and Pareto-based selection, offering a scalable alternative to RL-based methods such as GRPO \citep{shao2024deepseekmath}. Related reflection-driven approaches—PromptBreeder \citep{fernando2023promptbreeder}, EvoPrompt \citep{tong2025evoprompt}, AlphaEvolve \citep{novikov2025alphaevolve}, and TextGrad \citep{yuksekgonul2024textgrad}—further reinforce modular, sample-efficient PO ecosystems centered on reflective adaptation.
Beyond PO, Context Engineering (CE) \citep{mei2025context} generalizes optimization from prompts to entire context pipelines, encompassing retrieval, memory, and tool integration. While PO targets prompt design, CE focuses on system-level context orchestration—critical for robust, scalable, and production-grade LLMs capable of long-context reasoning.

\begin{figure*}[ht]
    \centering
    \includegraphics[width=0.83\linewidth]{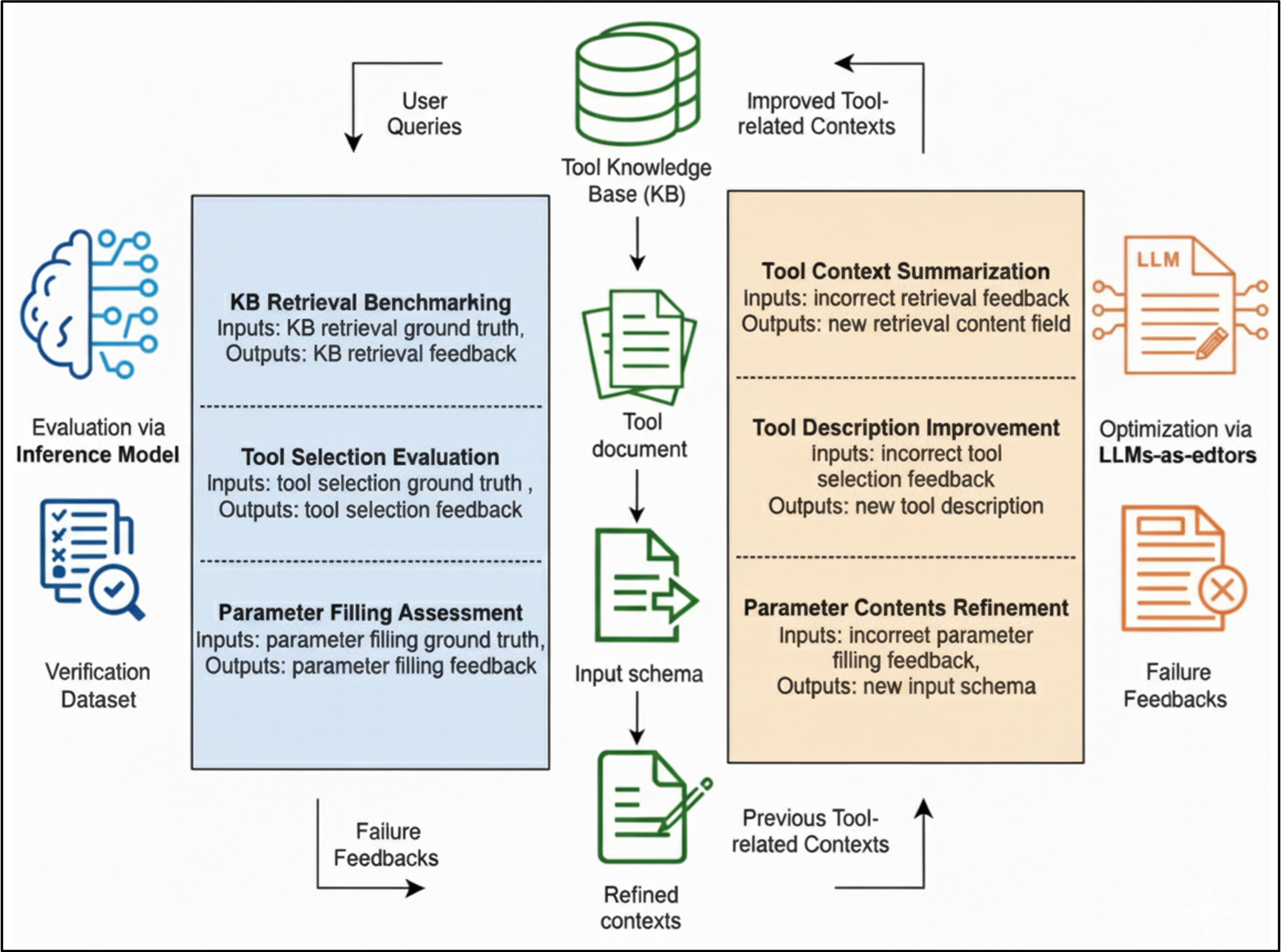}
    %\caption{\textbf{Verification-Guided Context Optimization (VGCO) Framework.} (1) User queries trigger evaluation through an inference model using a verification dataset. (2) The system conducts KB Retrieval Benchmarking, Tool Selection Evaluation, and Parameter Filling Assessment, each comparing inference outputs to ground-truth data and producing corresponding feedback. (3) These failure feedbacks identify incorrect retrievals, tool selections, or parameter fillings, which are passed to the refinement stage. (4) Using this feedback, LLMs-as-editors perform hierarchical context optimization across three sub-tasks: Tool Context Summarization (refining retrieval content fields), Tool Description Improvement (rewriting tool descriptions), and Parameter Content Refinement (updating input schemas). (5) The resulting refined contexts are integrated back into the Tool KB, forming a continuous feedback loop that iteratively improves tool-related context quality and inference accuracy over time.} 
    \caption{\textbf{Verification-Guided Context Optimization (VGCO) Framework.} 
    (1) User queries are evaluated via a verification dataset. (2) The system performs KB Retrieval, Tool Selection, and Parameter Filling assessments, producing feedback on errors. (3) LLMs-as-editors use this feedback to refine contexts through Tool Context Summarization, Tool Description Improvement, and Parameter Content Refinement. (4) The optimized contexts are reintegrated into the Tool KB, forming a feedback loop that improves context quality and inference performance.}
    \label{fig:framework}
\end{figure*}

\section{Preliminaries and Optimization Objective}
% In this section, we define the tool calling process in large language models (LLMs) and formulate the optimization objective for refining tool-related documentation and context.

\noindent\textbf{Tool Calling in LLMs.}
Tool calling refers to the process where a language model $\mathcal{M}$ generates and executes structured function calls to external tools in response to user queries. Given an input query $x \in X$ and tool-related contextual documentation $D$, the model produces a predicted tool call
\[
\tau = \mathcal{M}(D, x) = (\text{name}, \text{args}),
\]
where $\text{name} \in \mathcal{T} = \{t_1, t_2, \ldots, t_K\}$ represents the selected tool, and $\text{args} = \{(p_1, v_1), (p_2, v_2), \ldots, (p_j, v_j), \ldots\}$ denotes the set of parameter–value pairs required for executing the tool.
Each tool $t_i$ is described by a schema
\[
t_i = (\text{desc}_i, \text{params}_i),
\]
where $\text{desc}_i$ provides the textual description and $\text{params}_i$ specifies the valid parameter names, their data types, and whether each parameter is required or optional. The correctness of a tool call $\tau$ is determined by whether $\text{name}$ matches the intended tool and whether $\text{args}$ are consistent with the corresponding parameter schema.

\noindent\textbf{Tool Context Optimization.}
Given an initial tool-related documentation $D$, a language model $\mathcal{M}$, and a reward function $\mathcal{R}(\cdot)$ that measures the quality of model behavior (e.g., tool selection, parameter accuracy, or execution success), the objective is to learn an optimized documentation
\[
\widetilde{D} = \underset{D}{\arg\max} \; \mathbb{E}_{x \sim \mathcal{X}} \left[ \mathcal{R}(\mathcal{M}(D, x)) \right],
\]
where $D$ encompasses all contextual information relevant to tool calling—such as descriptions, schemas, relations, and usage examples—and $\mathcal{R}$ provides feedback signals reflecting task-level or behavioral performance. The refined documentation $\widetilde{D}$ maximizes the expected reward by improving the model’s ability to interpret queries, select appropriate tools, and generate correct and consistent tool calls under the optimized context.

\noindent\begin{definition}
\noindent \textbf{Hierarchically Tool Context Optimization.}
Let $V = (X, Y)$ denote a validation set with user queries $X = {x_1, \ldots, x_N}$ and corresponding ground-truth tool calls $Y = {y_1, \ldots, y_N}$. The documentation is structured hierarchically as $D = (D_r, D_t, D_p)$. Optimization proceeds top-down: first refining the retrieval context $D_r$, then the tool descriptions $D_t$, and finally the parameter schema $D_p$. Formally,
\[
\widetilde{D} = \underset{D}{\arg\max} \; eval(\mathcal{M}(D, X), Y),
\]
where $eval(\cdot)$ measures performance at the task level (e.g., retrieval recall, tool selection accuracy, or parameter completion score). The optimized $\widetilde{D}$ enhances retrieval relevance, tool selection reliability, and argument correctness.
\end{definition}

\noindent In this work, $D_r$ is an LLM-generated summarization of the full tool document, including tool description, input/output schemas, and system information, used exclusively for knowledge-base retrieval. $D_t$ consists of the tool descriptions used directly for tool selection. Finally, $D_p$ specifies the input schema, including parameter names, types, and required/optional roles, and serves as the instructions for parameter filling during tool invocation.
\section{Verification-Guided Context Optimization}

\subsection{Framework Overview}

To address the challenges of \textbf{scalability}, \textbf{variability}, and \textbf{ambiguity} in industrial-scale tool-calling, VGCO systematically refines hierarchical tool contexts—including retrieval fields, tool descriptions, and parameter schemas by combining structured verification signals with LLM-based editing. %As illustrated in Figure~\ref{fig:framework}, representative data are periodically collected from the online environment to capture evolving queries, tool usage, and error distributions while remaining decoupled from live inference. Offline evaluation of current contexts using metrics such as AST similarity, coverage, and intent alignment generates diagnostic signals to identify weak or outdated information. Guided by these signals, hierarchical LLM editors perform semantic-level refinements—additions, modifications, or deletions—across \textit{Tool Context Summarization}, \textit{Tool Description Improvement}, and \textit{Parameter Content Refinement}, with updates verified through rule-based checks and optional expert intervention. Refined contexts are validated offline to ensure stability and generalization before safe integration into the production environment. 
As shown in Figure~\ref{fig:framework}, representative data are periodically sampled to track evolving queries, tool usage, and errors. Offline diagnostics based on customized metrics guide hierarchical LLM editors to refine \textit{Tool Context Summarization}, \textit{Tool Description Improvement}, and \textit{Parameter Content Refinement}. Updates are verified and validated offline before integration into production.
Unlike self-refinement methods that rely solely on model-generated feedback, VGCO leverages explicit verification signals, ensuring interpretable, robust, and incremental improvements that adapt rapidly to real-world dynamics such as model updates, new tool onboarding, and shifting user domains without disrupting online inference or requiring retraining.

\begin{algorithm*}[ht]
\caption{VGCO: Hierarchical LLMs-as-Editors}
\label{alg:VGCO}
\begin{algorithmic}[1]
\STATE \textbf{Input:} Raw tool document $D = (D_r, D_t, D_p)$ with retrieval content fields $D_r$, tool descriptions $D_t$ and parameter contents $D_p$. Validation dataset $V=(X,Y)$, where $X$ is user query set and $Y$ is ground truth tool calling results (i.e, expected retrieval tool set, expected tool to be select and parameter to be fill); Inference model $\mathcal{M}$, retrieval editor $\mathcal{M}_r$, tool editor $\mathcal{M}_t$, parameter editor $\mathcal{M}_p$; Total iterations $T$.
\STATE \textbf{Output:} Optimized documentation $\widetilde{D} = (\widetilde{D}_r, \widetilde{D}_t, \widetilde{D}_p)$.
\STATE Initialize $\widetilde{D} \leftarrow D$.
\FOR{$t = 1$ to $T$}
    \FOR{each component $c \in \{r, t, p\}$ with editor $\mathcal{M}_c$}
        \STATE Collect negative examples $E_c$ via $\mathcal{M}(\widetilde{D}, \cdot)$ validated by $V$.
        \FOR{$e_c \in E_c$}
            \STATE Update $D_c \leftarrow \mathcal{M}_c(\widetilde{D}, e_c)$; $D \leftarrow (D_r, D_t, D_p)$.
            \IF{$eval(\mathcal{M}(D, X), Y) > eval(\mathcal{M}(\widetilde{D}, X), Y)$}
                \STATE $\widetilde{D} \leftarrow D$
            \ENDIF
        \ENDFOR
    \ENDFOR
\ENDFOR
\end{algorithmic}
\end{algorithm*}

\subsection{Hierarchical LLMs-as-Editors}

Tool-using contexts, including documentation and knowledge base contents, are naturally hierarchical, comprising retrieval fields, tool descriptions, and parameter schemas. Existing LLM-as-editor approaches \citep{qu2025exploration, yuan2024easytool} typically treat the entire document as a monolithic unit, neglecting the dependencies across contextual layers. To address challenges of granularity, dependency propagation, and targeted interpretability, we propose a \textbf{hierarchical LLM-as-editor} that performs structured, stage-wise optimization across multiple levels.

As illustrated in Figure~\ref{fig:framework} and detailed in Algorithm~\ref{alg:VGCO}, the framework refines tool contexts across three levels: retrieval ($D_r$), tool ($D_t$), and parameter ($D_p$). At each stage, failure cases identified through inference evaluation guide the corresponding editor modules ($\mathcal{M}_r, \mathcal{M}_t, \mathcal{M}_p$), and updates are retained only when post-edit evaluations demonstrate performance gains. The refinement proceeds hierarchically, ensuring that improvements at higher levels propagate effectively to lower levels.

\paragraph{Retrieval Level.}
The retrieval editor $\mathcal{M}_r$ refines $D_r$ using failed retrieval cases to improve semantic precision, coverage, and alignment with user queries. For example, in an API knowledge base where “Flight Status Query” and “Flight Route Search” are often confused, the editor rewrites retrieval contents by emphasizing distinguishing signals (e.g., temporal vs. spatial queries) and adding concrete examples. Refinements are validated via pre- and post-edit retrieval recall and precision metrics to ensure measurable gains in candidate identification.

\paragraph{Tool Level.}
The tool editor $\mathcal{M}_t$ revises $D_t$ when incorrect tool selections occur despite correct retrieval. For instance, when the model selects “BookHotelTool” instead of “BookFlightTool,” $\mathcal{M}_t$ analyzes mismatch logs, identifies overlapping intent phrases (“reservation,” “confirmation”), and rewrites tool descriptions with disambiguating cues and explicit positive/negative examples. Changes are retained only if verified to reduce misclassification in subsequent inference.

\paragraph{Parameter Level.}
The parameter editor $\mathcal{M}_p$ refines $D_p$ when parameter-filling errors persist after correct tool selection. For example, in “BookFlightTool,” repeated failures in filling the “DepartureDate” parameter (accepting text instead of ISO format) trigger schema-level edits clarifying data types and valid examples (e.g., “YYYY-MM-DD”). Revisions are validated through parameter accuracy metrics and schema conformance checks before being integrated.

This iterative process continues until predefined performance thresholds are met. By exploiting the hierarchical structure of tool contexts, the framework enables incremental, verifiable refinements, isolates the scope of edits, and minimizes unintended side effects, yielding a systematic mechanism for aligning documentation with observed LLM errors and improving overall tool-calling reliability. The rewritten examples of tool descriptions and parameter schemas are provided in Tables 6 and 7 in the Appendix.

\subsection{Editor with Guided Instruction}

Existing optimizers such as  Dspy \citep{khattab2023dspy}, SAMMO \citep{schnabel2024symbolic} and Promptim \citep{cui2025automatic} exhibit limited effectiveness in industrial-scale applications due to three main limitations: they operate in a \textit{stateless} manner, lack a \textit{well-defined action space}, and omit \textit{domain-specific reward signals}. To address these gaps, our framework introduces a \textbf{guided instruction editor} that performs \textit{state-aware}, \textit{action-constrained}, and \textit{domain-aligned} optimization across hierarchical levels.

Each editor operates under a structured \textit{guided instruction} paradigm, where its behavior is governed by a \textbf{guided system prompt} encoding four key elements:
\begin{enumerate}
    \item \textbf{State Space:} tracks prior edits and their performance outcomes, encompassing representative editing instances and the underlying reasoning processes, to prevent repetitive regressions and promote informed refinement.  
    \item \textbf{Action Space:} defines task-specific editing operations derived from successful revision patterns, guiding the model toward effective and targeted modifications (e.g., adding in-scope content, removing redundancy, unifying terminology, or abstracting hardcoded values).
    \item \textbf{Reward Signal:} validates errors query-by-query using ground truth labeling and inference reasoning, accompanied by reward signals to guide targeted refinement.
\end{enumerate}

\noindent In broader practice, the \textbf{state space} may include use-case domains (e.g., airline, retail, healthcare) and data sources from mixed datasets to enable context-aware refinement, while the \textbf{reward signal} can be applied batch-by-batch to balance effectiveness and efficiency, depending on the editor model’s context window and capabilities. In this work, the \textbf{action space} is customizable by optimization level: retrieval-level editors refine query–tool alignment; tool-level editors enhance semantic precision and domain relevance; and parameter-level editors refine schema definitions and examples to better align natural language with structured inputs. Each editor can embed shared background knowledge (e.g., tool conventions or syntax rules) to ensure coherence. Through structured prompts, adaptive actions, and domain-aligned evaluation, these guided editors enable targeted, verifiable optimization, addressing the stateless and unconstrained limitations of prior methods. The complete guided instruction editor prompts are presented in Tables 3-5 of the Appendix.

\section{Experiments}

\begin{figure*}[ht]
    \centering
    \begin{subfigure}{0.32\textwidth}
        \includegraphics[width=\linewidth]{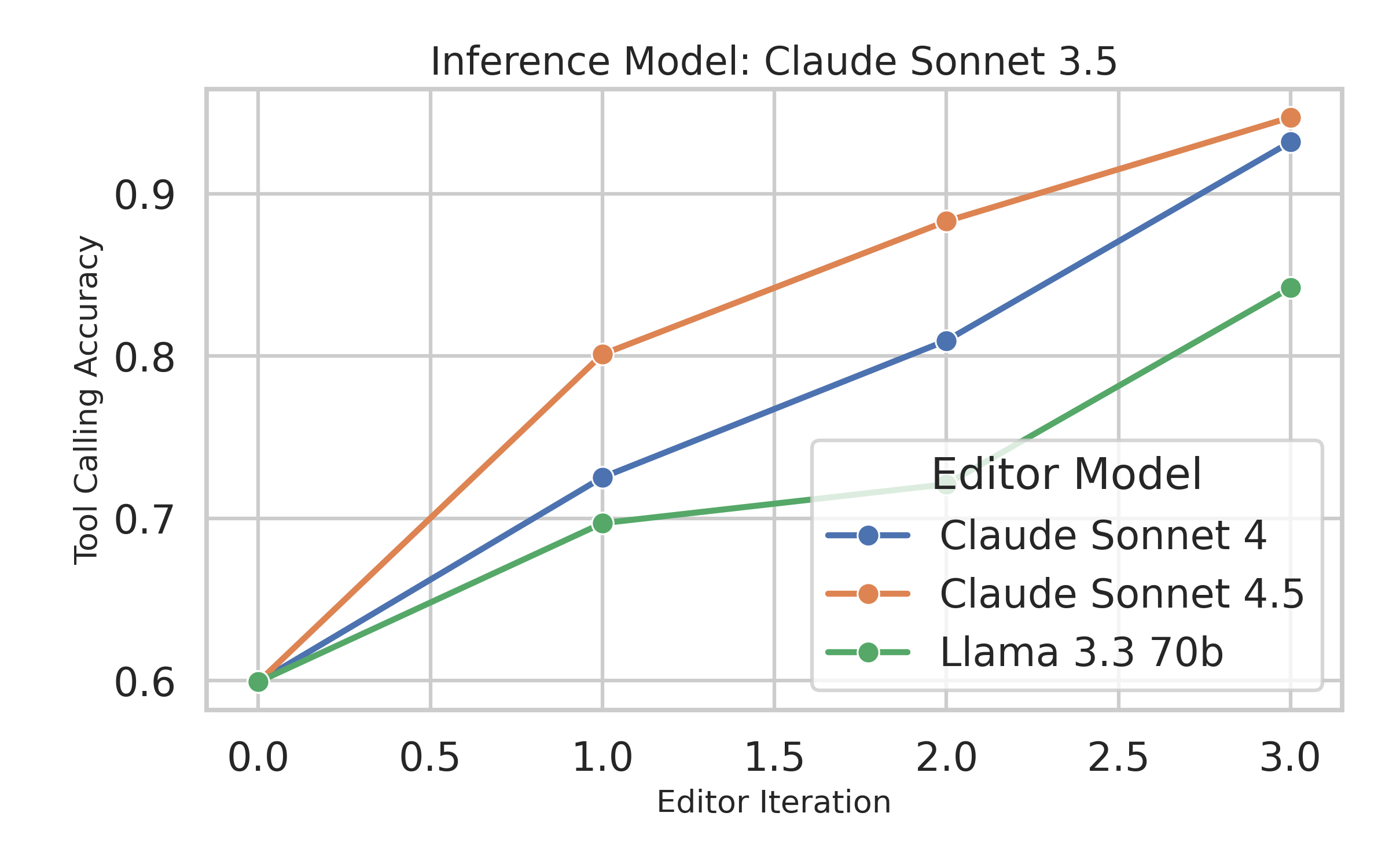}
        \caption{Claude Sonnet 3.5}
    \end{subfigure}
    \hfill
    \begin{subfigure}{0.32\textwidth}
        \includegraphics[width=\linewidth]{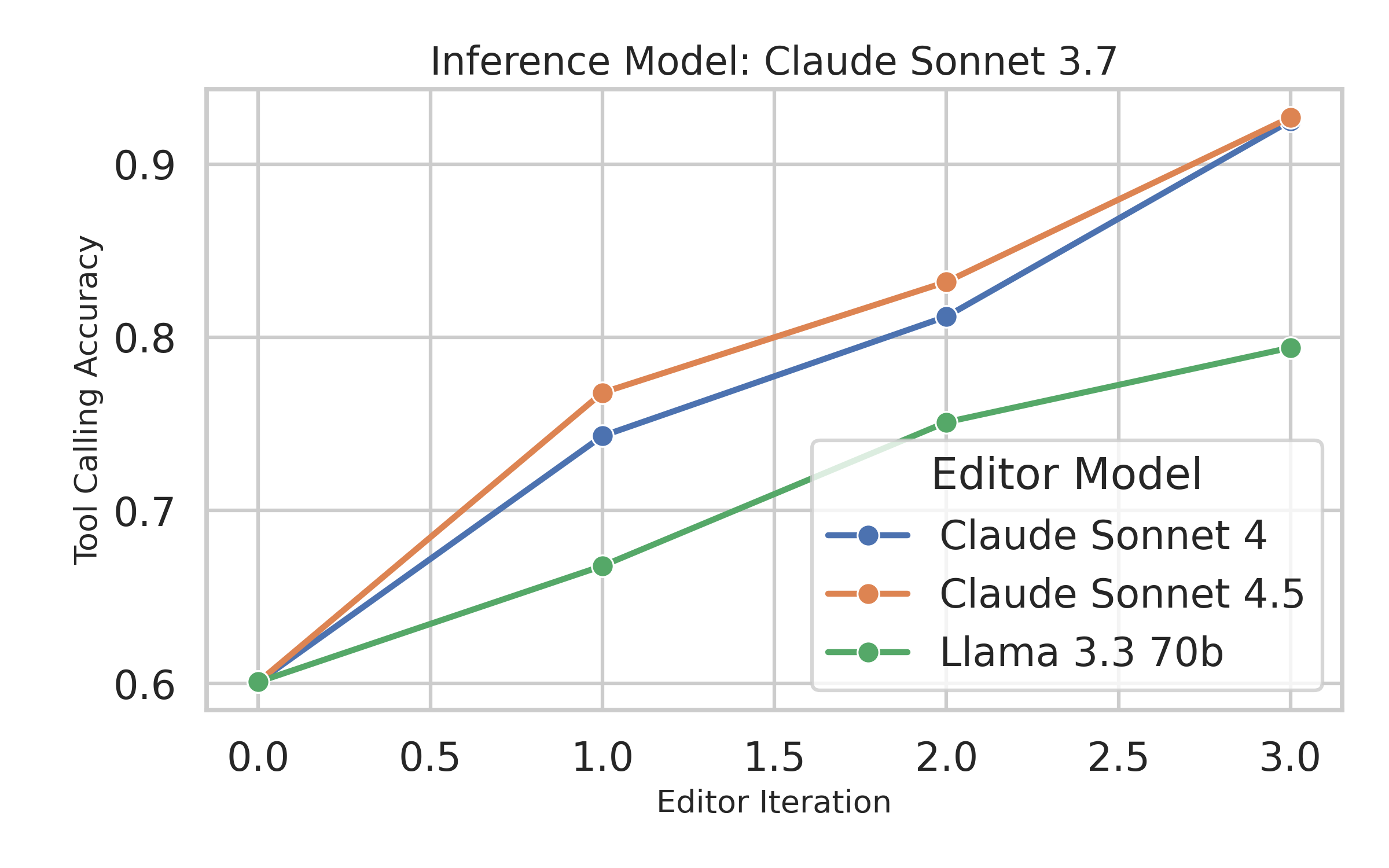}
        \caption{Claude Sonnet 3.7}
    \end{subfigure}
    \hfill
    \begin{subfigure}{0.32\textwidth}
        \includegraphics[width=\linewidth]{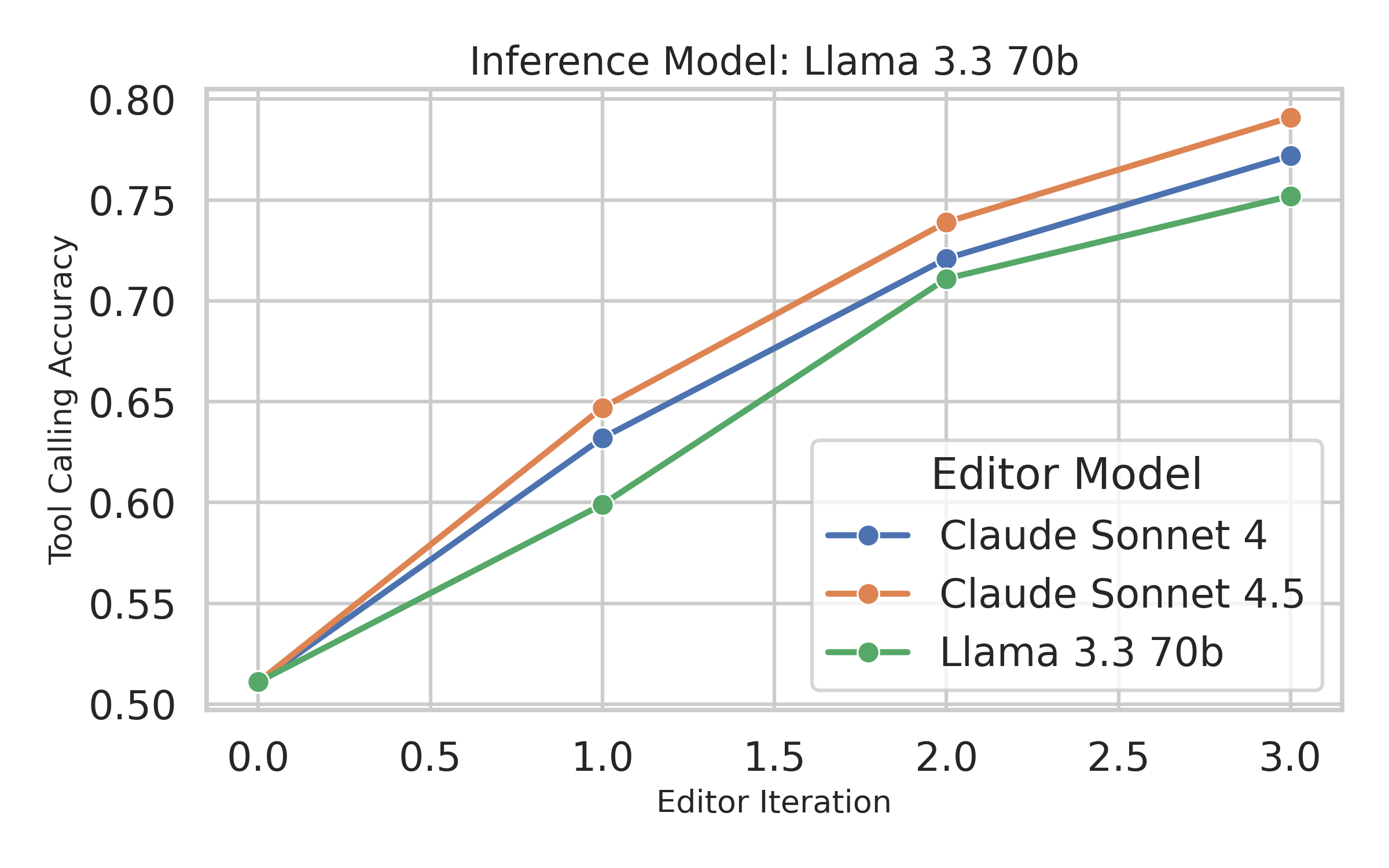}
        \caption{Llama 3.3 70b}
    \end{subfigure}
    \caption{Training accuracy comparison across editor iterations using Claude Sonnet 4, Claude Sonnet 4.5, and Llama 3.3 70B as editor models. Each editor is applied to Claude Sonnet 3.5, Claude Sonnet 3.7, and Llama 3.3 70B on the top-30 most common tools from the xLam dataset. To ensure fair comparison, training accuracy is evaluated on a subset of the training data whose tool-combination distribution matches that of the testing dataset.}
    \label{fig: training_accuracy_across_editor_models}
\end{figure*}

\begin{figure*}[ht]
    \centering
    \begin{subfigure}{0.32\textwidth}
        \includegraphics[width=\linewidth]{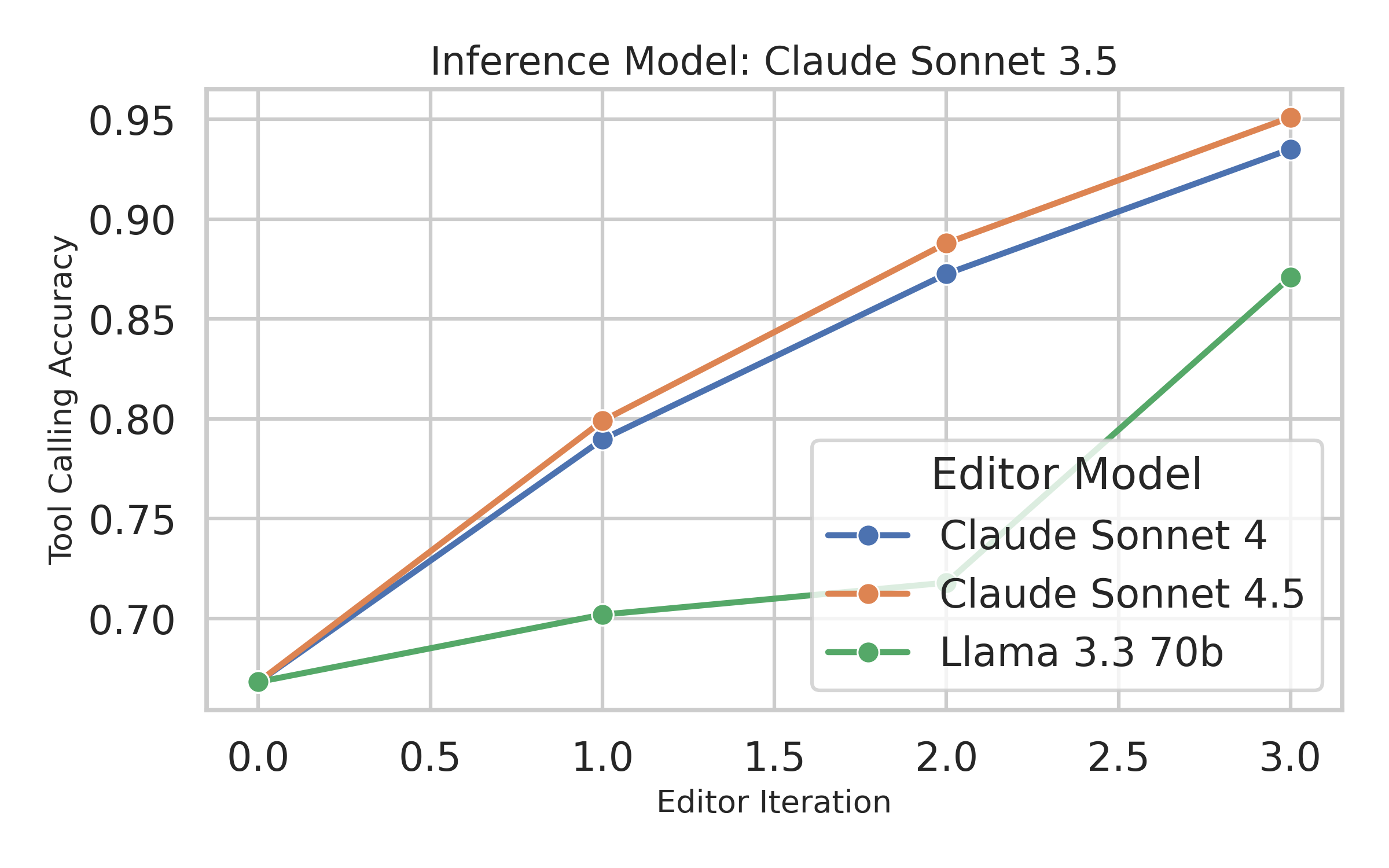}
        \caption{Claude Sonnet 3.5}
    \end{subfigure}
    \hfill
    \begin{subfigure}{0.32\textwidth}
        \includegraphics[width=\linewidth]{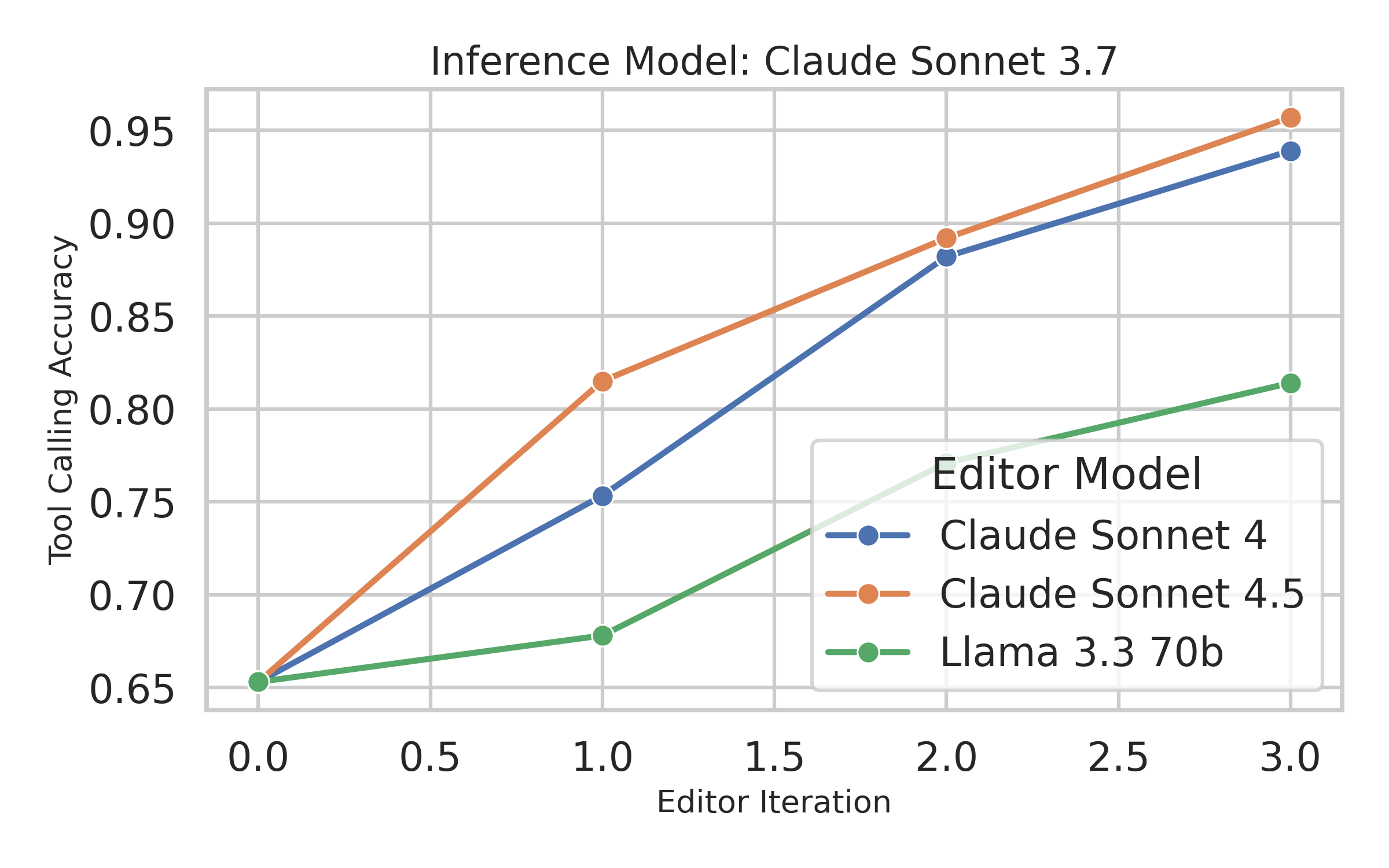}
        \caption{Claude Sonnet 3.7}
    \end{subfigure}
    \hfill
    \begin{subfigure}{0.32\textwidth}
        \includegraphics[width=\linewidth]{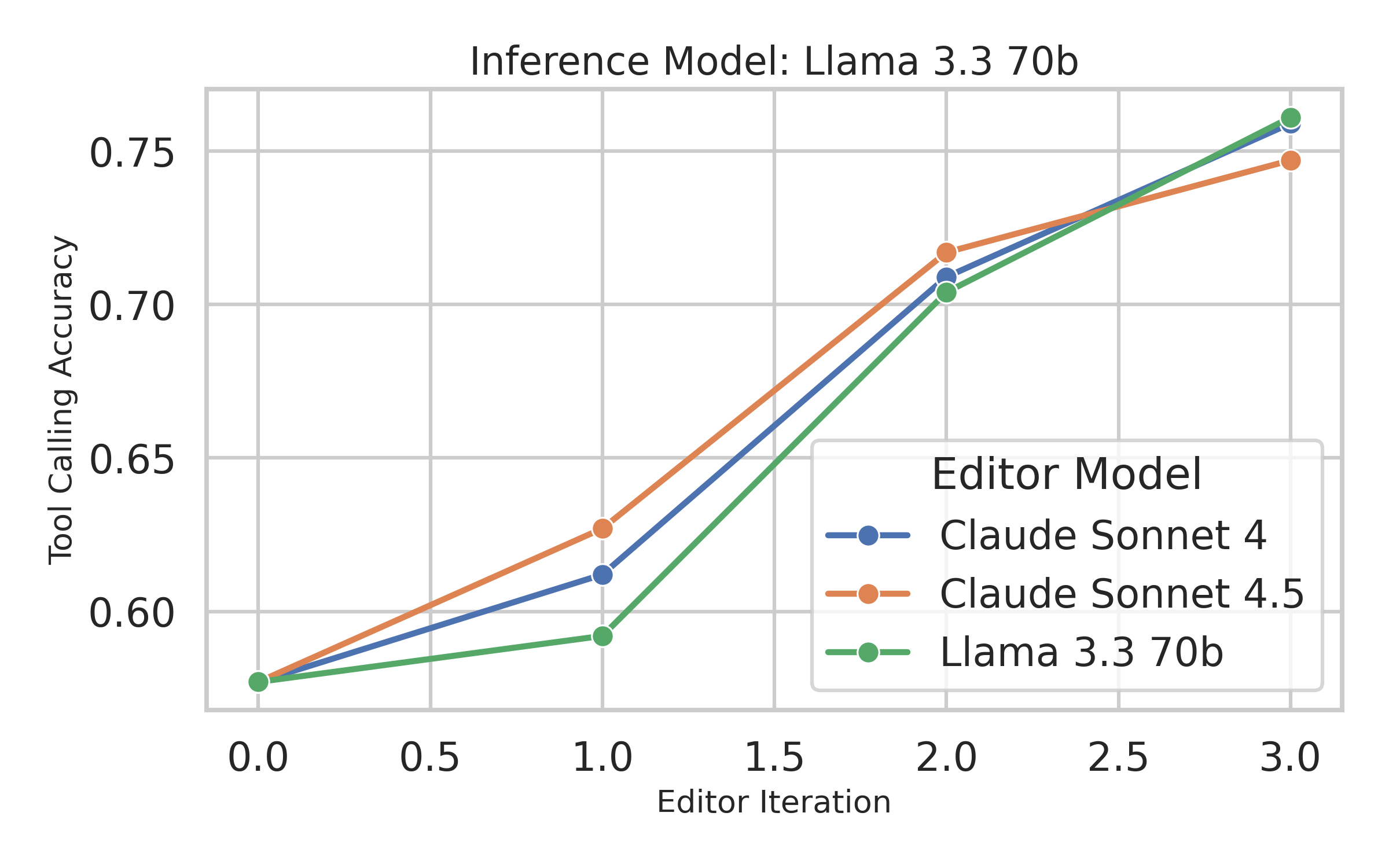}
        \caption{Llama 3.3 70b}
    \end{subfigure}
    \caption{Testing accuracy comparison across editor iterations using Claude Sonnet 4, Claude Sonnet 4.5, and Llama 3.3 70B as editor models. Each editor is applied to Claude Sonnet 3.5, Claude Sonnet 3.7, and Llama 3.3 70B on the top-30 most common tools from the xLam dataset.}
    \label{fig: testing_accuracy_across_editor_models}
\end{figure*}

\subsection{Datasets}
To simulate the large-scale tool invocation patterns characteristic of real-world industrial deployments, we evaluate our approach on two single-turn compositional benchmarks, xLAM \citep{zhang2024xlam} and BFCL \citep{patil2025bfcl}. xLAM requires parallel multiple tool calls per turn, where the ground truth is a list of tools that may include repeated calls but ignores the order of calls. The LLM is invoked only once to generate the response, operating in a single-turn, multi-choice setting. From the original 60k raw samples, we select the 100 most frequently used tools, extract their corresponding tool documentation, and compile them into a knowledge base to simulate large-scale tool usage in industrial settings. For each query, relevant tools are dynamically selected from this 100-tool set at initialization. At the lower level (tool selection and parameter filling), we focus on the top-10 tools discovered from the KB retrieval. The data are divided into training and testing sets based on unique answers (i.e., lists of tools): (1) if an entity contains only one query, it is assigned entirely to the training set; (2) if it contains multiple queries, they are approximately split into two-thirds for training and one-third for testing. All failure cases from model inference are revisited by the LLM editors using the training set, while the testing set remains strictly unseen during editing.
%BFCL-V1 focuses on the Parallel Multiple Function category, which combines multi-function selection and parallel execution, assessing the model’s ability to perform compositional reasoning across multiple function documents. We use the entire BFCL V1 dataset as the training set.
BFCL consists of three categories of function evaluation: %\textit{Simple Function}, 
\textit{Multiple Function}, \textit{Parallel Function}, and \textit{Parallel Multiple Function}. Simple Function evaluates single-function execution from one JSON document. Multiple Function tests the model’s ability to select the correct function from 2–4 candidates based on user context. Parallel Function requires invoking multiple function calls in response to a single query, which may span one or more sentences. Parallel Multiple Function combines these challenges, providing multiple function documents where each corresponding function call may be invoked zero or more times. In our experiments, we use all four BFCL datasets to comprehensively train and evaluate the model’s capability across function selection, parallel execution, and compositional reasoning.

%ToolBench is a large-scale benchmark comprising real-world APIs sourced from RapidAPI and BMTools, and is widely used to assess the tool-use capabilities of LLMs. Given budget constraints, our evaluation centers on the most challenging subset, I3-Instruction, which features complex user queries that require coordinating multiple tools across diverse categories. In addition, we utilize RestBench, a benchmark designed around two realistic application scenarios: TMDB, which includes 54 movie-related APIs, and Spotify, which provides 40 music-related APIs.

\subsection{Baselines}

%ReAct, Self-Reflexion based approach, DRAFT, Our LLM editors (with and without FT)
%(2) DFSDT\cite{qin2024toolllm}, which mitigates error propagation through a depth-first search strategy that improves decision-making accuracy; 
%(2) EasyTool, which simplifies tool descriptions by rewriting documentation with ChatGPT, enhancing LLMs’ understanding of tool functions and parameter usage;

We compare our \textbf{VGCO} framework against several widely used baselines. \textbf{Raw} represents the original tool documentation prior to any LLM editing. \textbf{ReAct}~\cite{yao2022react} combines reasoning and action, enabling the LLM to justify its steps and iteratively refine reasoning based on self-generated feedback from the environment. \textbf{DRAFT}~\cite{qu2025exploration} is a state-of-the-art dynamic framework that improves tool documentation via LLM-driven trial-and-error interactions, using a feedback-based process with phases for experience gathering, learning, and rewriting. In contrast, \textbf{VGCO} divides the tool-calling process into separate subtasks and addresses them independently—for example, performing tool selection only when retrieval quality is sufficient. While baseline methods may also apply iterative updates, they do not decompose the original tool-calling task into subtasks, instead editing the full task as a whole in each iteration.

\subsection{Evaluation Metrics}

% Abstract Syntax Tree (AST) for xLAM and BFCL
In the lower-level experiments, covering tool selection and parameter filling, on xLAM \citep{zhang2024xlam} and BFCL \citep{patil2025bfcl}, we implement an Abstract Syntax Tree (AST) validation that strictly enforces consistency between model outputs and function documentation. The AST checks function names, required parameters, and exact type matches; it also normalizes strings, interprets optional parameters contextually, and applies an all-or-nothing matching policy for multi-function outputs. For Knowledge Base tool retrieval, we evaluate performance using Recall@10, which measures whether all expected tools appear within the top-10 retrieved results. For tool selection and parameter filling tasks, we measure exact-match precision, meaning the predicted tool set must exactly match the ground truth, and all parameters must be filled with correct values.

\subsection{Main Results}

\begin{table*}[ht]
\centering
\small
\setlength{\tabcolsep}{8pt}
\renewcommand{\arraystretch}{1.15}
\begin{tabular}{l l c c c c}
\toprule
\textbf{Editor Model} & \textbf{Editing Method} & \textbf{Retrieval Recall} & \textbf{Tool Selection} & \textbf{Parameter Filling} & \textbf{Final Acc.} \\
\midrule
\multirow{4}{*}{Claude-4}
& Raw & 78.98 $\pm$ 0 & 71.5 $\pm$ 0.4 & 57.5 $\pm$ 0.3 & 32.5 \\
& ReAct & 91.64 $\pm$ 0.04 & 75.0 $\pm$ 0.6 & 56.0 $\pm$ 0.4 & 38.5 \\
& DRAFT & 95.62 $\pm$ 0.03 & 78.0 $\pm$ 0.7 & 62.0 $\pm$ 0.5 & 46.2 \\
& \textbf{VGCO (Ours)} & \textbf{97.48 $\pm$ 0.04} & \textbf{80.0 $\pm$ 0.5} & \textbf{65.0 $\pm$ 0.4} & \textbf{50.7} \\
\midrule
\multirow{4}{*}{Claude-4.5}
& Raw & 78.98 $\pm$ 0 & 71.5 $\pm$ 0.4 & 57.5 $\pm$ 0.3 & 32.5 \\
& ReAct & 93.02 $\pm$ 0.04 & 77.0 $\pm$ 0.4 & 62.0 $\pm$ 0.6 & 44.4 \\
& DRAFT & 96.59 $\pm$ 0.05 & 82.0 $\pm$ 0.5 & 71.0 $\pm$ 0.6 & 56.3 \\
& \textbf{VGCO (Ours)} & \textbf{98.30 $\pm$ 0.04} & \textbf{84.0 $\pm$ 0.3} & \textbf{74.0 $\pm$ 0.5} & \textbf{61.0} \\
\midrule
\multirow{4}{*}{Llama-3.3}
& Raw & 78.98 $\pm$ 0 & 71.5 $\pm$ 0.4 & 57.5 $\pm$ 0.3 & 32.5 \\
& ReAct & 87.42 $\pm$ 0.05 & 74.0 $\pm$ 0.4 & 60.0 $\pm$ 0.7 & 38.8 \\
& DRAFT & 94.07 $\pm$ 0.07 & 74.0 $\pm$ 0.5 & 62.0 $\pm$ 0.6 & 43.1 \\
& \textbf{VGCO (Ours)} & \textbf{97.24 $\pm$ 0.05} & \textbf{77.0 $\pm$ 0.6} & \textbf{65.0 $\pm$ 0.4} & \textbf{48.7} \\
\bottomrule
\end{tabular}
\caption{Performance on the top-100 most common tools from \textbf{xLam} dataset using Claude Sonnet 3.5 as the inference model. Reported metrics include tool retrieval recall (on the top-100 most commonly used tools), tool selection accuracy, parameter filling accuracy, and final execution accuracy (successful tool executions requiring correct retrieval, selection, and parameter filling). Each value represents the mean and standard deviation over 10 trials.}
\label{tab:xlam_results}
\end{table*}

\subsubsection{Effectiveness of LLM Editors.}

Figure~\ref{fig: training_accuracy_across_editor_models} and Figure~\ref{fig: testing_accuracy_across_editor_models} illustrate the progression of Tool Calling Accuracy across editor iterations (0–3) for different inference–editor configurations. Overall, accuracy consistently improves with additional editing cycles, confirming that VGCO’s iterative refinement effectively enhances tool-use precision and contextual reasoning. For the \textit{Claude Sonnet 3.5} inference model, the \textit{Claude Sonnet 4.5} editor achieves the highest final accuracies, reaching $\mathbf{0.947}$ (training) and $\mathbf{0.951}$ (testing) by Iteration~3. The \textit{Claude Sonnet 4} editor follows closely with $\mathbf{0.9318}$ (training) and $\mathbf{0.9351}$ (testing), while \textit{Llama-3.3-70b} trails behind at $\mathbf{0.842}$ (training) and $\mathbf{0.871}$ (testing). Similarly, for the \textit{Claude Sonnet 3.7} inference model, both \textit{Claude Sonnet 4.5} and \textit{Claude Sonnet 4} converge near the top with $\mathbf{0.927}$ and $\mathbf{0.925}$ in training, and $\mathbf{0.957}$ and $\mathbf{0.939}$ in testing, respectively. The \textit{Llama-3.3-70b} editor remains lower, ending at $\mathbf{0.794}$ (training) and $\mathbf{0.814}$ (testing). When the inference model is \textit{Llama-3.3-70b}, performance across editors is lower overall, with the \textit{Claude Sonnet 4.5} editor again performing best, reaching $\mathbf{0.791}$ (training) and $\mathbf{0.747}$ (testing), closely followed by \textit{Claude Sonnet 4} at $\mathbf{0.772}$ (training) and $\mathbf{0.759}$ (testing), and the \textit{Llama-3.3-70b} editor itself at $\mathbf{0.752}$ (training) and $\mathbf{0.761}$ (testing). Across all models, the largest gains occur in the first two iterations (e.g., from $0.5992$ to $0.801$ training accuracy for \textit{Claude Sonnet 3.5} with \textit{Claude Sonnet 4.5}), followed by gradual convergence as contextual alignment stabilizes. These results confirm that VGCO’s hierarchical editing mechanism functions as a structured self-improvement process, with \textit{Claude Sonnet 4.5} consistently delivering the strongest and most stable improvements across inference models and datasets.

\subsubsection{Comparison with Baseline Methods.}

Table~\ref{tab:xlam_results} presents a detailed comparison between the proposed \textbf{VGCO} framework and representative baselines—ReAct, DRAFT, and the unedited Raw variant—evaluated across multiple inference models. VGCO consistently achieves the highest performance across all dimensions, including retrieval recall, tool selection accuracy, parameter filling, and the composite final accuracy obtained from their joint product.
Across all model backbones, VGCO delivers steady and significant gains. For instance, with \textit{Claude-4}, VGCO attains a final accuracy of 50.7\%, surpassing DRAFT by over 4 points and ReAct by more than 12 points. Under \textit{Claude-4.5}, the improvement grows to 61.0\%, alongside the highest retrieval recall (98.3\%) and parameter filling (74.0\%). Comparable trends are observed with \textit{Llama-3.3}, indicating that VGCO generalizes well across model architectures without requiring specific tuning.
VGCO’s advantages are most pronounced in tool selection and parameter filling—subtasks that demand fine-grained contextual reasoning. These gains stem from VGCO’s hierarchical, guided editing process, which constrains each stage to targeted improvements such as ambiguity reduction, redundancy elimination, and context enrichment. In contrast, baseline methods like ReAct and DRAFT rely on heuristic or stateless operations, leading to less stable performance and greater variance.
Overall, VGCO demonstrates that structured, multi-level editing substantially enhances reasoning alignment and compositional precision, offering a robust and scalable solution for context optimization in tool-using LLM systems.

\subsection{Additional Experiments}

\subsubsection{BFCL Results.}

Table~\ref{tab:bfcl_claude4} presents the comparative results of different editing methods on the \textbf{BFCL} dataset, evaluated across two dimensions: \textit{Tool Selection} and \textit{Parameter Filling}. The baseline model (\textit{Raw}) achieves moderate performance with accuracies of $\mathbf{69.1\%}$ and $\mathbf{55.8\%}$, respectively. Incorporating reasoning-based correction via \textit{ReAct} improves results to $\mathbf{85.0\%}$ in Tool Selection and $\mathbf{74.4\%}$ in Parameter Filling, reflecting gains from step-by-step deliberation. The instruction-editing approach \textit{DRAFT} further elevates accuracy to $\mathbf{92.1\%}$ and $\mathbf{88.3\%}$, indicating effective alignment through structured edit representation. However, the proposed \textbf{VGCO} framework achieves the highest performance, reaching $\mathbf{96.9\%}$ in Tool Selection and $\mathbf{94.7\%}$ in Parameter Filling—surpassing all baselines by a clear margin. These results confirm VGCO’s superior ability to refine tool-use reasoning and parameter precision through iterative, guided contextual optimization, establishing it as the most effective editing strategy on the BFCL benchmark.

\begin{table}[H]
\centering
\small
\setlength{\tabcolsep}{6pt}
\renewcommand{\arraystretch}{1.15}
\begin{tabular}{l c c c c}
\toprule
\textbf{Editing Method} & \textbf{Tool Selection} & \textbf{Parameter Filling}\\
\midrule
Raw & 69.1 $\pm$ 0.2 & 55.8 $\pm$ 0.3\\
ReAct & 85.0 $\pm$ 0.2 & 74.4 $\pm$ 0.4\\
DRAFT & 92.1 $\pm$ 0.3 & 88.3 $\pm$ 0.5\\
\textbf{VGCO (Ours)} & \textbf{96.9 $\pm$ 0.1} & \textbf{94.7 $\pm$ 0.2}\\
\bottomrule
\end{tabular}
\caption{Performance comparison of editing methods using the Claude Sonnet 4 model as editor and Claude Sonnet 3.7 model as inference on the \textbf{BFCL} dataset.}
\label{tab:bfcl_claude4}
\end{table}

\subsubsection{Ablation Studies of Guided Instruction.}

\begin{figure}[H]
    \centering
    \includegraphics[width=\linewidth]{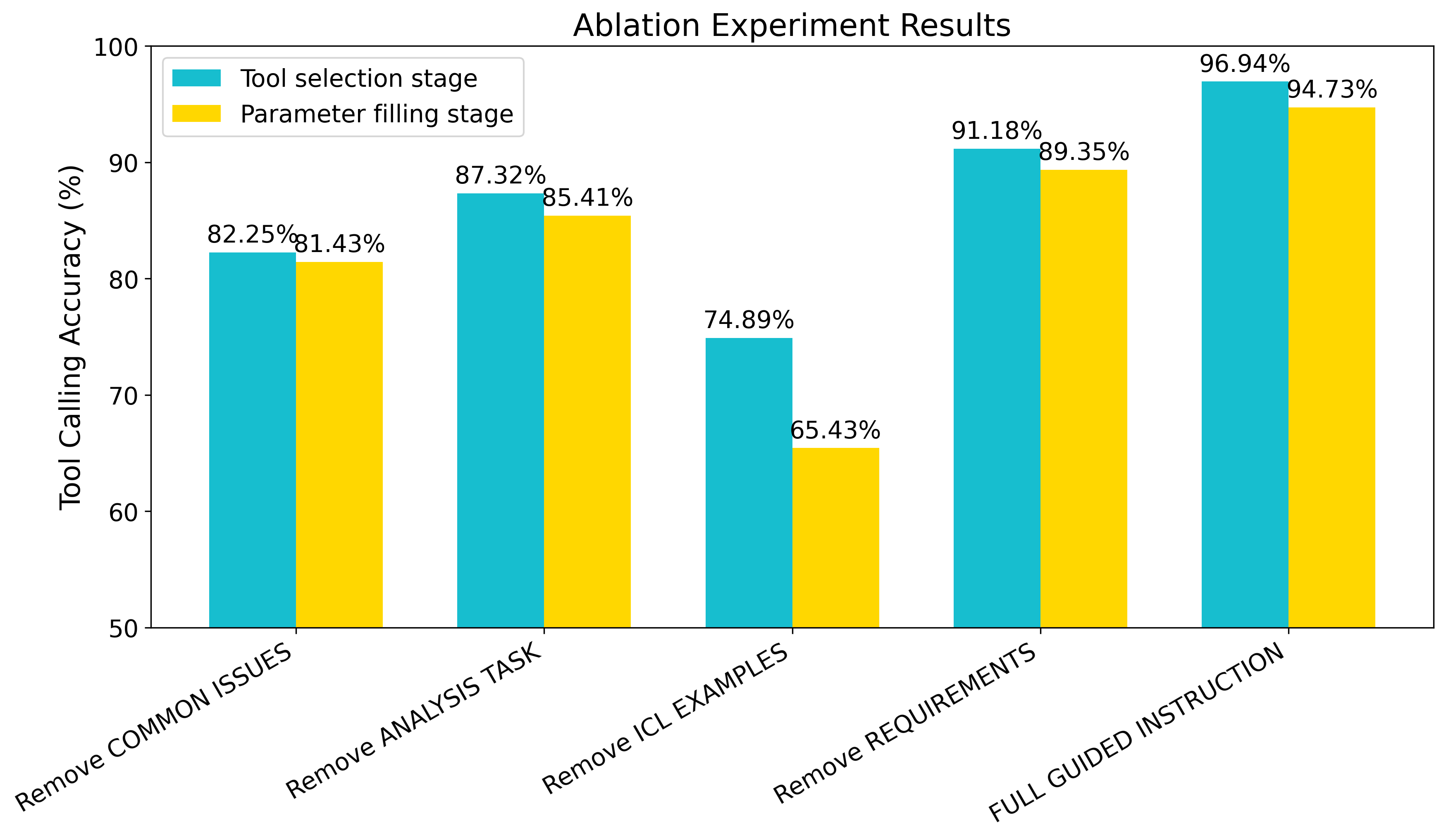}
    \caption{Ablation studies of each component in \textbf{GUIDED INSTRUCTION} using the Claude Sonnet 4 model as editor and Claude Sonnet 3.7 model as inference  on the \textbf{BFCL} dataset. FULL GUIDED INSTRUCTION is provided in Tables 3-5 of Appendix.}
    \label{fig:ablation_results}
\end{figure}

Figure~\ref{fig:ablation_results} presents the results of ablation experiments that assess the contribution of each component in the guided instruction used by the VGCO editors. The evaluation compares Tool Calling Accuracy across two stages—\textit{Tool Selection} and \textit{Parameter Filling}—under progressively reduced instruction configurations. Removing the \textbf{Common Issues} section, which enumerates frequent tool misuse patterns, results in a moderate drop to $\mathbf{82.25\%}$ (selection) and $\mathbf{81.43\%}$ (filling), showing that explicit error typologies improve corrective precision. Excluding the \textbf{Analysis Task} (responsible for structured mismatch reasoning) slightly decreases performance to $\mathbf{87.32\%}$ and $\mathbf{85.41\%}$, respectively, indicating that analytical framing supports stable model focus but is not the dominant driver. Omitting \textbf{In-Context Learning (ICL) Examples} reduces accuracy to $\mathbf{74.89\%}$ and $\mathbf{65.43\%}$, highlighting their strong influence on contextual alignment and adaptation. Removing \textbf{Requirements}, which enforce clarity and modification constraints, leads to a sharper decline to $\mathbf{91.18\%}$ and $\mathbf{89.35\%}$, confirming the importance of structural editing guidance for coherence and consistency. Finally, the \textbf{Full Guided Instruction} configuration yields the highest accuracies—$\mathbf{96.94\%}$ in Tool Selection and $\mathbf{94.73\%}$ in Parameter Filling—demonstrating the cumulative benefit of combining analytical reasoning, issue typologies, structured requirements, and ICL examples. Collectively, these results validate that the guided instruction design in VGCO provides essential scaffolding for interpretable, self-corrective improvement in both selection and parameterization stages.

% \noindent 1. generalizability to different tool set (top-100)/model

% \noindent 2. unit editor test on * BFCL (level and action, ICL)

% \noindent 3. * start from zero document (* SP-API?)

% \noindent 4. * FT experiments

% SFT and DPO on small Qwen models compared with Qwen w/o post-training

\section{Conclusion}

% \textbf{limitation:} scalability and robustness, potentially addressed by Tool Pre-Selection and Inference Model Post-training

% \noindent \textbf{broader impact:} failure handler, hallucination, and all context engineering can help with

% \noindent \textbf{future works:} 1. more context to optimize, e.g., user query, response, state (temporal degree), knowledge-based retrieval, personal context, etc. 2. online RL (PPO, GRPO) on LLMs-as-editors (eval metrics -> reward functions) 3. complex task like multi-turn 3. enhance inference model + CO to achieve better performance 4. tool-sing dataset with real interactions like AppWorld

In this work, we presented Verification-Guided Context Optimization (VGCO), a unified framework that leverages LLMs-as-editors to iteratively refine tool documentation and knowledge base contexts for improved tool use. By tightly integrating evaluation feedback, reasoning traces, and structured online-offline learning, VGCO demonstrates strong empirical improvements in both tool-use accuracy and generalization across diverse models and domains.

\subsubsection{Limitation: } While VGCO substantially alleviates scalability challenges by maintaining effective performance across hundreds of tools, further experiments are required to validate its generalizability in even larger and more heterogeneous tool ecosystems. Moreover, although VGCO is designed to be model-agnostic, its robustness under domain shift, unseen tools, or low-resource scenarios requires additional validation. These limitations may be addressed by incorporating tool pre-selection strategies to further reduce context size or by employing post-training adaptation methods tailored for tool-augmented inference environments.

\subsubsection{Broader Impact and Future Work:} 
The VGCO framework advances the development of safer and more reliable LLM-based systems by embedding verification-guided context optimization into the learning loop. This design not only mitigates tool misuse and hallucination under unfamiliar or ambiguous conditions but also enables self-corrective behavior through verification-based feedback. Looking forward, extending VGCO beyond tool descriptions and KB contents to include user queries, system prompts, interaction histories, and personalized context may further enhance adaptability and task relevance. Incorporating online reinforcement learning methods such as PPO \citep{schulman2017proximal} or GRPO \citep{shao2024deepseekmath} can establish a more principled, scalable loop for continual editor improvement. Future efforts will also explore multi-turn conversational interaction scenarios requiring temporal reasoning and memory, as well as joint optimization between inference-time models and the context optimizer to strengthen their synergy. Finally, benchmarking VGCO on realistic tool-use datasets, e.g., AppWorld \citep{trivedi2024appworld}, will be essential for validating robustness and generalization in production-like environments.

\bibliography{aaai2026}

%\end{multicols}

\onecolumn

\newpage

\section{Appendix}

\begin{multicols}{2}

\subsection{LLM Editors Post-training}

%Post-training LLM editors offers two key benefits for enhancing the VGCO framework. First, it enables online or in-context self-play, where editors can be optimized through either SFT, DPO with both positive and negative editing ICL examples or reinforcement learning (e.g., PPO or GRPO) using evaluation-derived reward signals, allowing continual improvement in aligning context edits with tool-use success. Second, it supports knowledge distillation from large, high-performing models to smaller, more efficient models such as those in the same model families (e.g., Qwen or LLaMA), enabling cost-effective deployment while preserving editing capabilities. Together, these strategies enhance both the adaptability and scalability of LLM-as-editor systems.
Post-training LLM editors provide two major advantages for enhancing the VGCO framework. First, they enable online or in-context self-play, where editors are continually optimized via SFT, DPO with both positive and negative in-context editing examples, or reinforcement learning methods (e.g., PPO, GRPO) guided by evaluation-derived reward signals. This process continually improves the alignment between context edits and successful tool use. Second, post-training facilitates knowledge distillation from larger, high-performing models to smaller, more efficient ones within the same family (e.g., Qwen, LLaMA), enabling cost-effective deployment while retaining editing proficiency. Together, these strategies enhance the adaptability, efficiency, and scalability of LLM-as-editor systems.

\subsubsection{Supervised Fine-Tuning}

Supervised Fine-Tuning (SFT) is the standard approach to align language models with task-specific datasets through maximum likelihood estimation (MLE). Given a dataset $\mathcal{D}_{\text{sft}} = \{(x_i, y_i)\}_{i=1}^N$ of input-output pairs, SFT minimizes the negative log-likelihood (NLL) of the target output $y_i$ given input $x_i$:

\begin{equation}
\mathcal{L}_{\text{SFT}}(\theta) = -\frac{1}{N} \sum_{i=1}^{N} \log \pi_\theta(y_i \mid x_i),
\end{equation}

\noindent where $\pi_\theta$ is the model parameterized by $\theta$.

\subsubsection{Direct Preference Optimization}

Direct Preference Optimization (DPO) is a method designed to train models from human preferences without requiring reinforcement learning. Given a dataset $\mathcal{D}_{\text{pref}} = \{(x_i, y_i^+, y_i^-)\}_{i=1}^N$, where $y_i^+$ is the preferred response over $y_i^-$ for prompt $x_i$, DPO maximizes the log-likelihood ratio between the preferred and dispreferred responses under the target policy $\pi_\theta$ and a fixed reference policy $\pi_{\text{ref}}$. The DPO loss is defined as:

% \begin{equation}
% \mathcal{L}_{\text{DPO}}(\theta) = -\frac{1}{N} \sum_{i=1}^{N} \log \sigma\left( \beta \left[ \log \frac{\pi_\theta(y_i^+ \mid x_i)}{\pi_{\text{ref}}(y_i^+ \mid x_i)} - \log \frac{\pi_\theta(y_i^- \mid x_i)}{\pi_{\text{ref}}(y_i^- \mid x_i)} \right] \right),
% \end{equation}

\begin{align}
\mathcal{L}_{\text{DPO}}(\theta) = -\frac{1}{N} \sum_{i=1}^{N} \log \sigma\Bigg( \beta \Big[ \log \frac{\pi_\theta(y_i^+ \mid x_i)}{\pi_{\text{ref}}(y_i^+ \mid x_i)} \nonumber \\
\quad - \log \frac{\pi_\theta(y_i^- \mid x_i)}{\pi_{\text{ref}}(y_i^- \mid x_i)} \Big] \Bigg),
\end{align}

\noindent where $\sigma(z) = \frac{1}{1 + e^{-z}}$ is the sigmoid function, and $\beta > 0$ is a temperature parameter controlling the sharpness of the preference.

Intuitively, DPO encourages the model to increase the relative likelihood of preferred outputs over dis-preferred ones while staying close to the reference policy $\pi_{\text{ref}}$. 

\subsubsection{Group Relative Preference Optimization}

Group Relative Preference Optimization (GRPO) extends preference-based learning by comparing model outputs across groups of responses rather than isolated pairs, enabling more stable and fine-grained alignment. Given a dataset $\mathcal{D}_{\text{grp}} = \{(x_i, \mathcal{Y}_i)\}_{i=1}^N$, where $\mathcal{Y}_i = \{y_{i,1}, \ldots, y_{i,m}\}$ represents a group of $m$ candidate responses for prompt $x_i$, each response $y_{i,j}$ is associated with a scalar preference score $r_{i,j}$ (e.g., human rating, reward model output, or heuristic score).

GRPO optimizes the relative log-likelihood of each response within its group, encouraging the model to assign higher probabilities to higher-rated outputs. The objective is defined as:

\begin{equation}
\mathcal{L}_{\text{GRPO}}(\theta) = -\frac{1}{N} \sum_{i=1}^{N} \sum_{j=1}^{m} w_{i,j} \log \pi_\theta(y_{i,j} \mid x_i),
\end{equation}

\noindent where the normalized weight $w_{i,j}$ reflects the relative preference of response $y_{i,j}$ within the group, typically computed as

\begin{equation}
w_{i,j} = \frac{\exp(\beta r_{i,j})}{\sum_{k=1}^{m} \exp(\beta r_{i,k})},
\end{equation}

with $\beta$ controlling the sensitivity to preference differences.

Intuitively, GRPO generalizes DPO by leveraging groupwise comparisons rather than binary preferences, improving sample efficiency and robustness when multiple graded or partially ordered responses are available. This formulation allows smooth preference propagation across multiple candidates and facilitates joint optimization under diverse evaluation signals.

\subsection{Implementation Details}

\paragraph{Knowledge Base and Storage Configuration.}
For knowledge retrieval, we employ the Amazon Bedrock Knowledge Base to implement a Retrieval-Augmented Generation (RAG) pipeline for the Tool Knowledge Base (ToolKB). The system uses Amazon OpenSearch Serverless as the vector store and the cohere-embed-english-v3 model to generate 1024-dimensional float embeddings. Tool definitions are stored as JSON schemas specifying each tool’s name, purpose, and input parameters; these structured descriptions are embedded and indexed for semantic retrieval. During ingestion, a hierarchical hybrid chunking strategy (300-token chunks with 20-token overlap) maintains semantic coherence and retrieval precision. At query time, a hybrid search approach integrates semantic vector matching with lexical search, following the Bedrock RAG workflow to embed queries, retrieve semantically aligned chunks, and augment model prompts for accurate and context-aware tool retrieval. All experiment data and results are stored in Amazon S3 using a general-purpose bucket configuration that supports mixed access patterns and multi–Availability Zone redundancy.

\paragraph{Model and Throttling Configuration.}
All experiments are conducted using Amazon Bedrock, a managed service that provides secure access to high-performance foundation models through a unified API. Bedrock’s serverless architecture enables efficient deployment and execution without manual infrastructure management. In our experimental setup, Bedrock serves as the unified inference backend for both retrieval and editing stages, ensuring consistent configuration and reproducibility across experiments.
For inference evaluation, we employ Claude~3.5~Sonnet, Claude~3.7~Sonnet, and Llama~3.3~70B~Instruct as the primary base models. Each model is run with deterministic settings (\texttt{temperature=0.0}, \texttt{topP=1.0}) and maximum token limits of 4096 for Claude models and 2048 for Llama. These models are selected for cost efficiency, as inference involves over 10,000 query examples per iteration to compute comprehensive metrics across training and testing datasets.
For iterative refinement in VGCO, we utilize higher-performance editor models, including Claude~4~Sonnet, Claude~4.5~Sonnet, and Llama~3.3~70B~Instruct. Editor parameters mirror the base configuration to maintain stability across editing iterations. High-capacity editor models are preferred because they process only a subset of examples—primarily failure cases from prior iterations—while handling richer contextual information, such as inference reasoning and in-context examples, within the model’s context window.
To mitigate throttling during large-scale evaluation, we adopt a custom policy that limits concurrency (\texttt{max\_requests=2}) and request rate (\texttt{5~req/min}), with up to three retries per request using exponential backoff. This configuration provides a balanced trade-off between robustness and throughput in long-running inference experiments.

\begin{table*}[ht]
\centering
\small
\renewcommand{\arraystretch}{1.1}
\caption{\textbf{Tool Retrieval Editor System Prompt.}}
\begin{tabular}{p{0.96\textwidth}}
\toprule
\textbf{Retrieval Optimization Editor} \\ 
\midrule
\textbf{Task Description:} \\
You are an expert in optimizing knowledge base tool retrieval contents for retrieval-augmented language models. 
Your task is to analyze retrieval mismatches and improve KB retrieval contents to prevent retrieval errors. \\[4pt]

\textbf{Input Data:} \\
Current KB Retrieval Contents: \texttt{\{KB\_retrieval\_contents\}} \\
Retrieval Mismatch Examples (query, expected\_tools, retrieved\_tools, retrieval\_response): \texttt{\{retrieval\_mismatches\}} \\[4pt]

\textbf{Analysis Task:} \\
Analyze each retrieval mismatch example. Identify patterns in why the KB retrieves or selects incorrect tools. Determine which KB tool contents require refinement. Focus on clarity, specificity, and distinguishing information that improves retrieval recall rate. \\[4pt]

\textbf{Common Retrieval Issues:}
\begin{itemize}
\item Ambiguous or overlapping KB tool descriptions
\item Missing concrete use cases or negative examples
\item Unclear distinctions between similar tools
\item Incomplete parameter or context details
\item Insufficient guidance on when and why to use each tool
\end{itemize}

\textbf{Output Format:} \\
\texttt{ANALYSIS} — Detailed analysis of retrieval issues and reasoning for improvements. \\
\texttt{IMPROVED KB TOOL DESCRIPTIONS} — Updated KB retrieval contents in JSON format with headers: ‘name’, ‘retrieval content’. \\[4pt]

\textbf{Requirements:}
\begin{itemize}
\item Only modify retrieval contents for tools in mismatch examples
\item Keep all other retrieval contents unchanged
\item Add examples and clarify distinctions
\item Maintain retrieval consistency with prior optimizations
\end{itemize}
\\[4pt]
\textbf{In-Context Learning Examples:} \texttt{\{retrieval\_icl\_examples\}} \\[4pt]
\bottomrule
\end{tabular}
\label{tab:retrieval_prompt}
\end{table*}

\begin{table*}[ht]
\centering
\small
\renewcommand{\arraystretch}{1.1}
\caption{\textbf{Tool Selection Editor System Prompt.}}
\begin{tabular}{p{0.96\textwidth}}
\toprule
\textbf{Tool Selection Optimization Editor} \\ 
\midrule
\textbf{Task Description:} \\
You are an expert in optimizing tool descriptions for tool-use language models. 
Your task is to analyze tool mismatches and improve tool descriptions to prevent these errors. \\[4pt]

\textbf{Input Data:} \\
Current Tool Descriptions: \texttt{\{tool\_use\_descriptions\}} \\
Tool Mismatch Examples (query, expected\_tools, actual\_tools, model\_response): \texttt{\{tool\_mismatches\}} \\[4pt]

\textbf{Analysis Task:} \\
Analyze each mismatch example. Identify why the model selects wrong tools and determine which descriptions require improvement. \\[4pt]

\textbf{Common Tool Selection Issues:}
\begin{itemize}
\item Ambiguous tool descriptions
\item Missing key use cases or examples
\item Overlapping functionality between tools
\item Unclear parameter requirements
\item Missing context about when to use each tool
\end{itemize}

\textbf{Output Format:} \\
\texttt{ANALYSIS} — Detailed analysis of tool selection errors and rationale for each improvement. \\
\texttt{IMPROVED TOOL DESCRIPTIONS} — Updated tool descriptions in JSON format with headers: ‘name’, ‘description’. \\[4pt]

\textbf{Requirements:}
\begin{itemize}
\item Modify only tools appearing in mismatch examples
\item Maintain clarity, specificity, and distinguishing features
\item Include examples for complex tools
\end{itemize}
\\[4pt]
\textbf{In-Context Learning Examples:} \texttt{\{tool\_icl\_examples\}} \\[4pt]
\bottomrule
\end{tabular}
\label{tab:tool_prompt}
\end{table*}

\begin{table*}[ht]
\centering
\small
\renewcommand{\arraystretch}{1.1}
\caption{\textbf{Parameter Filling Editor System Prompt.}}
\begin{tabular}{p{0.96\textwidth}}
\toprule
\textbf{Parameter Filling Optimization Editor} \\ 
\midrule
\textbf{Task Description:} \\
You are an expert in optimizing parameter descriptions to improve parameter filling accuracy in tool-use language models. 
Your task is to analyze parameter mismatches and enhance input schema clarity. \\[4pt]

\textbf{Input Data:} \\
Current Tool Input Schemas: \texttt{\{tool\_input\_schemas\}} \\
Parameter Mismatch Examples (query, param\_coverage\_ratio, param\_all\_match, tools\_schema\_expected, tools\_schema\_actual): \texttt{\{parameter\_mismatches\}} \\[4pt]

\textbf{Analysis Task:} \\
Analyze each parameter mismatch, identify the cause, and refine schema descriptions and examples. \\[4pt]

\textbf{Common Parameter Issues:}
\begin{itemize}
\item Incorrect parameter types or formats
\item Missing required parameters
\item Misunderstood parameter meanings
\item Confusion between similar parameters
\end{itemize}

\textbf{Output Format:} \\
\texttt{ANALYSIS} — Explanation of parameter filling errors and reasoning for schema changes. \\
\texttt{IMPROVED TOOL DESCRIPTIONS} — Updated tool schemas in JSON format with headers: ‘name’, ‘tools’. \\[4pt]

\textbf{Requirements:}
\begin{itemize}
\item Modify only mismatched parameter schemas
\item Clarify parameter purpose, type, and required status
\item Add usage examples and specify expected formats
\end{itemize}
\\[4pt]
\textbf{In-Context Learning Examples:} \texttt{\{parameter\_icl\_examples\}} \\[4pt]
\bottomrule
\end{tabular}
\label{tab:parameter_prompt}
\end{table*}

\begin{table*}[ht]
\centering
\small
\setlength{\tabcolsep}{6pt}
\renewcommand{\arraystretch}{1.2}
\begin{tabular}{p{0.95\textwidth}}
\toprule
\textbf{VGCO Modifications for Tool Selection} \\
\midrule

\textbf{Get Product Tool (\textcolor{red}{single-product-per-call})} \\
\textbf{Raw:} \{"Tool Name": "get\_product", "description": "Retrieves product information including name, price, and specifications for a specific product ID."\} \\
\textbf{Ours:} \{"Tool Name": "get\_product", "description": "Retrieves product information for a single product ID per call including name, price, and specifications."\} \\
\textbf{Example Queries:} 
\begin{itemize}
    \item "Get the product information for the item with ID 101112, but if it fails, try with ID 131415."
    \item "Retrieve details for the product with ID 12345. Also, fetch details for the product with ID 67890."
\end{itemize} \\

\midrule

\textbf{Find Pairs with Sum Tool (\textcolor{red}{target sum not emphasized})} \\
\textbf{Raw:} \{"Tool Name": "find\_pairs\_with\_sum", "description": "Identifies all pairs of numbers within an array that add up to a specified target sum value."\} \\
\textbf{Ours:} \{"Tool Name": "find\_pairs\_with\_sum", "description": "Finds all pairs of numbers in an array that add up to a specified target sum."\} \\
\textbf{Example Queries:} 
\begin{itemize}
    \item "Find all pairs of integers in the list [1, 2, 3, 4, 5, 6, 7, 8] that sum up to 12."
\end{itemize} \\

\midrule

\textbf{Chi-Square Independence Test (\textcolor{red}{statistical relationship})} \\
\textbf{Raw:} \{"Tool Name": "chi\_square\_independence\_test", "description": "Performs Chi-Square test for independence between two categorical variables using contingency table data."\} \\
\textbf{Ours:} \{"Tool Name": "chi\_square\_independence\_test", "description": "Tests statistical independence between two categorical variables using frequency data to determine if variables are related."\} \\
\textbf{Example Queries:} 
\begin{itemize}
    \item "Compute a Chi-Square test for independence on a table where 80 people like action movies and 20 like romance, with 60 being male and 40 being female."
\end{itemize} \\

\midrule

\textbf{Bacterial Growth Tool (\textcolor{red}{multi-calculation-per-call})} \\
\textbf{Raw:} \{"Tool Name": "bacterial\_growth", "description": "Calculates bacterial population size for one growth scenario using initial count, growth rate, and time parameters. Processes one calculation per call."\} \\
\textbf{Ours:} \{"Tool Name": "bacterial\_growth", "description": "Calculates bacterial population size for a single set of parameters including initial count, growth rate, and time requiring separate calls for each calculation."\} \\
\textbf{Example Queries:} 
\begin{itemize}
    \item "Given an initial population of 1500 bacteria and a growth rate of 0.04 per minute, predict the population after 35 minutes. Also, consider the case where the doubling time is 30 minutes."
    \item "Calculate the bacterial population after 30 minutes if the initial population is 1500 and the growth rate is 0.5. Also, find the population after 50 minutes with the same initial population but a growth rate of 0.7."
\end{itemize} \\

\midrule

\textbf{Structural Analysis Tool (\textcolor{red}{specific analysis details})} \\
\textbf{Raw:} \{"Tool Name": "structural\_analysis", "description": "Performs engineering structural analysis on buildings using building ID, floor numbers, and analysis type parameters."\} \\
\textbf{Ours:} \{"Tool Name": "structural\_analysis", "description": "Performs engineering structural analysis including dynamic, static, and seismic analysis on building structures using building ID and floor specifications."\} \\
\textbf{Example Queries:} 
\begin{itemize}
    \item "What city is associated with zip 30303? Also, do a dynamic analysis on building 789 for floors 2, 4, and 6."
    \item "Identify the city for ZIP code 30318. Carry out a dynamic structural analysis on building 1011 for floors 4, 5, and 6."
\end{itemize} \\

\bottomrule
\end{tabular}
\caption{Tool description rewriten examples showing the original and VGCO-modified description along with example queries highlighting mismatch issues.}
\label{tab:vgco_tool_updates}
\end{table*}

\begin{table*}[ht]
\centering
\small
\setlength{\tabcolsep}{6pt}
\renewcommand{\arraystretch}{1.2}
\begin{tabular}{p{0.95\textwidth}}
\toprule
\textbf{VGCO Modifications for Parameter Filling} \\
\midrule

\textbf{Power-of-Two Check Tool (\textcolor{red}{parameter schema creation})} \\
\textbf{Raw:} \{"Tool Name": "is\_power\_of\_two", "parameters": []\} \\
\textbf{Ours:} \{"Tool Name": "is\_power\_of\_two", "parameters": [\{"name": "numbers", "description": "A single number or list of numbers to check if they are powers of two. Can be integers like 8, 16, 1024, or a list like [8192, 16384]. Required parameter.", "type": "array|integer"\}]\} \\
\textbf{Example Queries:} 
\begin{itemize}
    \item "For a memory allocation task, I need to know if 8192 and 16384 are powers of two."
\end{itemize} \\

\midrule

\textbf{IP Geolocation Tool (\textcolor{red}{batch and zip inclusion support})} \\
\textbf{Raw:} \{"Tool Name": "ip\_geolocation", "parameters": []\} \\
\textbf{Ours:} \{"Tool Name": "ip\_geolocation", "parameters": [\{"name": "ip\_address", "description": "A single IP address to geolocate and retrieve ZIP code information for.", "type": "string"\}, \{"name": "include\_zip\_code", "description": "Whether to include ZIP/postal code in the geolocation response.", "type": "boolean", "default": "true"\}, \{"name": "batch\_mode", "description": "Enable batch processing for multiple IP addresses.", "type": "boolean", "default": "false"\}]\} \\
\textbf{Example Queries:} 
\begin{itemize}
    \item "I have a list of IP addresses and I need to find their corresponding ZIP codes: 172.217.16.110, 172.217.16.100, 172.217.16.102."
\end{itemize} \\

\midrule

\textbf{Sum-Pair and Growth Tools (\textcolor{red}{multi-function schema generation})} \\
\textbf{Raw:} \{"Tool Name": "find\_sum\_pairs", "parameters": []\}, \{"Tool Name": "predict\_bacterial\_growth", "parameters": []\} \\
\textbf{Ours:} \{"find\_sum\_pairs": [\{"name": "numbers", "description": "Array of numbers to search for pairs.", "type": "array"\}, \{"name": "target\_sum", "description": "The target sum value that pairs should equal.", "type": "number"\}], "predict\_bacterial\_growth": [\{"name": "initial\_population", "description": "Initial bacterial population count at time zero.", "type": "number"\}, \{"name": "growth\_rate", "description": "Bacterial growth rate constant (per unit time).", "type": "number"\}, \{"name": "time\_minutes", "description": "Time duration in minutes for which to predict the population.", "type": "number"\}]\} \\
\textbf{Example Queries:} 
\begin{itemize}
    \item "Find pairs in [5, 7, 9, 11, 13] that sum to 12, and predict the bacterial count after 20 minutes with an initial population of 1100 and a growth rate of 0.04."
\end{itemize} \\

\midrule

\textbf{Electric Field Calculation Tool (\textcolor{red}{added medium permittivity parameter})} \\
\textbf{Raw:} \{"Tool Name": "calculate\_electric\_field", "parameters": [\{"charge": "int", "distance": "int"\}]\} \\
\textbf{Ours:} \{"Tool Name": "calculate\_electric\_field", "parameters": [\{"name": "charge", "description": "The electric charge in Coulombs (required). Example: 5", "type": "number"\}, \{"name": "distance", "description": "The distance from the charge in meters (required). Example: 1.5", "type": "number"\}, \{"name": "permittivity", "description": "Permittivity of the medium (optional). Default: 8.854e-12 F/m", "type": "number", "default": 8.854e-12\}]\} \\
\textbf{Example Queries:} 
\begin{itemize}
    \item "Calculate the electric field produced by a 5 Coulomb charge at a distance of 1 meter, and then at 2 meters."
\end{itemize} \\

\midrule

\textbf{ZIP and Structural Analysis Tools (\textcolor{red}{split into independent parameter sets})} \\
\textbf{Raw:} \{"Tool Name": "zip\_code\_lookup", "parameters": []\}, \{"Tool Name": "dynamic\_structural\_analysis", "parameters": []\} \\
\textbf{Ours:} \{"zip\_code\_lookup": [\{"name": "zip\_code", "description": "Required 5-digit ZIP code to identify the city location.", "type": "string"\}], "dynamic\_structural\_analysis": [\{"name": "building\_id", "description": "Unique identifier for the building to analyze.", "type": "string"\}, \{"name": "floors", "description": "List of floor numbers to include in the structural analysis.", "type": "array"\}]\} \\
\textbf{Example Queries:} 
\begin{itemize}
    \item "Identify the city for ZIP code 30318. Carry out a dynamic structural analysis on building 1011 for floors 4, 5, and 6."
\end{itemize} \\

\midrule

\textbf{WHOIS Domain Tool (\textcolor{red}{improved type and validation})} \\
\textbf{Raw:} \{"Tool Name": "whois", "parameters": [\{"domain": "str"\}]\} \\
\textbf{Ours:} \{"Tool Name": "whois", "parameters": [\{"name": "domain", "description": "The domain name to lookup WHOIS information for. Must be a valid domain (e.g., 'google.com').", "type": "string"\}]\} \\
\textbf{Example Queries:} 
\begin{itemize}
    \item "Get information about the domain 'yahoo.com' and 'microsoft.com' using the WHOIS function."
\end{itemize} \\

\bottomrule
\end{tabular}
\caption{Parameter schema rewritten examples. Each case illustrates schema completion or correction based on missing or ambiguous parameter definitions inferred from natural-language queries.}
\label{tab:vgco_parameter_updates}
\end{table*}

\end{multicols}

\end{document}